\documentclass[10pt,letterpaper]{article}

\usepackage{ccn}
\usepackage{pslatex}
\usepackage{apacite}
\usepackage[utf8]{inputenc} %
\usepackage[T1]{fontenc}    %
\usepackage{url}            %
\usepackage{booktabs}       %
\usepackage{amsfonts}       %
\usepackage{nicefrac}       %
\usepackage{microtype}      %

\usepackage{tikz}
\usetikzlibrary{arrows}

\usepackage{amsmath}
\usepackage{algorithm2e}
\usepackage{mathtools,enumitem}
\usepackage[title]{appendix}
\usepackage{graphicx,import}
\usepackage{subcaption}
\usepackage{mwe}
\graphicspath{ {./figures/} }
\usepackage{tabularx}
\usepackage{xcolor}
\newcommand{\Real}{\mathbb{R}}

\newcommand{\qed}{\nobreak \ifvmode \relax \else
      \ifdim\lastskip<1.5em \hskip-\lastskip
      \hskip1.5em plus0em minus0.5em \fi \nobreak
      \vrule height0.75em width0.5em depth0.25em\fi}

\newcommand{\set}[1]{\{#1\}}
\newcommand{\beq}{\begin{equation}}
\newcommand{\eeq}{\end{equation}}

\newcommand{\tran}{^{\text{\sf T}}}

\title{Low-Rank Nonlinear Decoding of $\mu$-ECoG from the Primary Auditory Cortex \thanks{This work was presented at 2018 \textit{Conference on Cognitive Computational Neuroscience (CCN)}, Philadelphia, Pennsylvania.}}

\author{
{\large \bf Melikasadat Emami$^1$, Mojtaba Sahraee Ardakan$^1$, Parthe Pandit$^1$, 
Alyson K. Fletcher$^2$} \\
{\large \{emami,msahraee,parthepandit,akfletcher\}@ucla.edu } \\
  $^1$Department of Electrical and Computer Engineering, $^2$Department of Statistics, University of California, Los Angeles,\\ 
  420 Westwood Plaza, Los Angeles, CA 90095
  \AND {\large \bf Sundeep Rangan,} srangan@nyu.edu \\
  Department of Electrical and Computer Engineering, New York University,\\ 
  2 MetroTech Center, Brooklyn, NY 11201
 \AND {\large \bf Michael Trumpis, Brinnae Bent, Chia-Han Chiang, Jonathan Viventi}\\
 \{michael.trumpis,brinnae.bent,ken.chiang,j.viventi\}@duke.edu  \\
  Department of Biomedical Engineering, Duke University\\
101 Science Drive Durham, NC 27708}

\begin{document}

\maketitle

\section{Abstract}
{
\bf
This paper considers the problem of neural decoding from
parallel neural measurements systems such as micro-electrocorticography ($\mu$-ECoG).
In systems with large numbers of array elements at
very high sampling rates, the dimension of the raw measurement data may be large.
Learning neural decoders for this high-dimensional data can be challenging,
particularly when the number of training samples is limited.  
To address this challenge, 
this work presents a novel neural network decoder with a low-rank
structure in the first hidden layer.  The low-rank constraints dramatically
reduce the number of parameters in the decoder while
still enabling a rich class of nonlinear decoder maps.
The low-rank decoder is illustrated on $\mu$-ECoG data from the 
primary auditory cortex (A1) of awake rats. This decoding problem
is particularly challenging due to the complexity of neural responses in the
auditory cortex and the presence of confounding signals in awake
animals.  It is shown that the proposed low-rank decoder significantly outperforms
models using standard dimensionality reduction techniques such as 
principal component analysis (PCA).

}
\begin{quote}
\small
\textbf{Keywords:}
auditory decoding; neural networks; low-rank filter; dimensionality reduction
\end{quote}  %

\section{Introduction}

Advancements in neural recording technologies, particularly calcium imaging and high-dimensional
micro-electrocorticography ($\mu$-ECoG), now enable  measurements of tremendous
numbers of neurons or brain regions in parallel \cite{chang2015towards,fukushima2015studying,stosiek2003vivo}.
While these recordings offer the potential to
observe neural activity at unprecedented level
of detail, the high-dimensionality presents a fundamental challenge for learning
neural decoding systems from data.

This dimensionality problem is particularly acute for the focus of this work,
namely neural decoding of signals in the primary auditory
signals from state-of-the-art $\mu$-ECoG.
Most importantly, in modern $\mu$-ECoG systems,
the dimensionality of the meaured responses often exceeds the
number of training examples.  For example, in the application we discuss below
the responses from the $\mu$-ECoG array \cite{insanally2016low}
for each stimuli consists of approximately 160 time samples across 61 electrodes,
resulting in a raw feature dimension of $(160)(61) = 9760$.  However,
due to experimental limits on the duration of the experiments,
there are less than 400 training examples.  Moreover,
responses in the primary
auditory cortex are known to be complex \cite{zatorre2002structure, mlynarski2018learning}.
Also, for awake animals, the responses may have confounding components from movements.
Consequently, neural decoding systems must be sufficiently rich
to enable nonlinear decoding and confounding signal rejection.

This work presents a novel approach for neural decoding from parallel
neural measurements with a small number of parameters
while being able to capture complex nonlinear relationships between the measurements
and stimulus.  The approach is based on a traditional neural network structure,
but with two key novel properties:
(1) A discrete-cosine transform (DCT) pre-processing stage
used to reduce the sampling rate; and
(2) An initial linear layer of the neural network with a low rank structure.
We argue that both structures are well-justified based on the physical
processes and can dramatically reduce the number of parameters.
The method is demonstrated in neural decoding in the rat primary
auditory cortex (A1) from a new high-dimensional $\mu$-ECoG array \cite{insanally2016low}.

\subsection{Previous work}
Despite advancements in machine learning tools,
traditional methods are still common in auditory decoding~\cite{glaser2017machine}.
Some of these methods consider both linear and non-linear mapping of the neural responses to the
auditory spectrogram \cite{pasley2012reconstructing}. Linear neural decoders like support vector machines (SVM) have also been widely discussed to classify behavioral responses using population activity \cite{francis2018small}. As in \cite{de2018decoding} other methods like canonical correlation analysis (CCA) have also been used as linear models to measure the correlation between the stimulus and response as a goodness of fit after transforming them. 

Multi-layer neural networks showed remarkable success in feature extraction and classification in machine vision and speech processing \cite{yamins2016using}. Since auditory signals arrive to the cortex after having been passed through a number of sensory processing areas, these networks are appealing to model the responses in auditory cortex \cite{hackett2011information}. 

There is also a large body of literature in dimensionality reduction methods
for high-dimensional neural recordings \cite{cunningham2014dimensionality,mazzucato2016stimuli,williamson2016scaling, sadtler2014neural}.  
The methods are largely based on the unlabeled data and attempt to find
a low-dimensional latent representation that can capture the bulk of the signal variance.
Neural decoders can then be trained on the low-dimensional representation to reduce
the number of parameters.
As we will see in the results section below, our method
can outperform these dimensionality reduction-based techniques since
the proposed method operates on the labeled data and, in essence, find the directions of
variance that are best tuned for the neural decoding task.

\section{Model Description}

We consider the problem of decoding stimuli from
$d$-dimensional neural responses recorded from some area of the brain.
Such responses can arise from any parallel measurements system including
responses measured by an ECoG microelectrode array with $d$ channels,
calcium traces from $d$ neurons, or signals recorded by the recently developed Neuropixel probes \cite{callaway2017brain}.
Let $X^i\in \Real^{d\times T}$ be the response to some stimulus $y^i$ recorded in a time window of length $T$ after the stimulus is applied. Given $N$ input-output sample pairs
$\set{(X^1,y^1), \dots, (X^N,y^N)}$,
the neural decoding problem is to learn a decoder that can estimate the stimulus
$y$ from a new response $X$. Depending on whether the stimuli $y$ is discrete or continuous-valued,
the decoding problem can be viewed either as a classification or regression.

The key challenge in this decoding problem is the potential high-dimensionality of the
input to the decoder $X$.
Since the response $X$ has $p=dT$ features, even linear classification or regression
would require $O(dT)$ parameters.  This number of parameters may easily exceed
the number of trials $N$ on which the decoder can be trained. Thus, some form
of dimensionality reduction or structure on the decoder is required.

\begin{figure}[t]\centering
\scalebox{0.25}{
\begin{tikzpicture}[thick,->,draw=black!80, node distance=2.5cm, scale=1, every node/.style={scale=1.5, font=\large}]
\tikzstyle{every pin edge}=[<-,line width=3pt]
\tikzstyle{neuron}=[circle,fill=black!25,minimum size=30pt,inner sep=0pt]

\tikzstyle{lowrank neuron}=[circle,fill=black!25,minimum size=55pt,inner sep=0pt, fill=yellow!20];

\tikzstyle{hidden neuron}=[neuron, fill=red!30];

\tikzstyle{annot} = [text width=1em, text centered]
\tikzstyle{null} = [draw = none , fill= none]
\tikzstyle{line} = [draw, black, -latex']
\tikzstyle{X block}=[rectangle,draw=blue!65,fill=blue!25,minimum size=40pt, minimum height=3cm,minimum width=4cm]

\tikzstyle{Z block}=[rectangle,draw=green!65, fill=green!35,minimum size=40pt, minimum height=3cm,minimum width=1.5cm]

\tikzstyle{V}=[rectangle,fill=cyan!25,minimum size=4pt, minimum height=1cm,minimum width=2mm]

\tikzstyle{U}=[rectangle,fill=black!25,minimum size=4pt, minimum height=2mm,minimum width=1cm]

\node[rectangle, fill = black!5, draw=black!20, minimum width = 13.5cm , minimum height= 9cm, rounded corners=5mm, xshift = 9.5cm, yshift = 0.25cm](model){};

\node[annot, above of = model, yshift = 2.5cm, text width = 10cm, font=\LARGE]{Low-Rank Decoder};

\node[X block, font=\LARGE](X){$\mathbf{X}$};
\node[annot, above of = X, yshift = -0.7cm, xshift = -0.3cm, font=\LARGE](input){input};
\node[annot, below of = X, yshift = 0.7cm, xshift = -0.3cm, font=\LARGE]{time};
\node[annot, left of = X, xshift = 0.2cm,yshift = 0cm, font=\LARGE]{d};

\node[Z block, right of= X, xshift = 3cm, font=\LARGE](Z){$\mathbf{Z}_0$};
\node[annot, below of = Z, yshift = 0.7cm, font=\LARGE]{freq};
\node[annot, left of = Z, yshift = 0cm, xshift = 1.5cm, font=\LARGE](d){d};

\path (X) edge [above, line width=0.8mm, draw=black!50] node[text width = 4em, xshift = 0.5cm, font=\LARGE]{DCT}(d);

\node [lowrank neuron, right of = Z, yshift = 2.9cm, xshift = 1cm, font=\LARGE] (l1){$\quad \mathbf{v}_1 \mathbf{u}_1\tran$};
\node[V, right of=Z, xshift = 0.5cm, yshift = 3cm, font=\LARGE](v1){};
\node[U, right of=Z, xshift = 1.2cm, yshift = 3.4cm, font=\LARGE](u1){};

\node [lowrank neuron, below of = l1, yshift = 0.3cm, font=\LARGE] (l2) {$\quad \mathbf{v}_2 \mathbf{u}_2\tran$};
\node[V, right of=Z, xshift = 0.5cm, yshift = 0.8cm, font=\LARGE](v2){};
\node[U, right of=Z, xshift = 1.2cm, yshift = 1.2cm, font=\LARGE](u2){};

\node[annot, below of = l2, yshift = 0.7cm ]{$\vdots$};

\node [lowrank neuron, below of = l2, yshift = -1cm, font=\LARGE] (l3) {$\quad \mathbf{v}_3 \mathbf{u}_3\tran$};
\node[V, right of=Z, xshift = 0.5cm, yshift = -2.7cm, font=\LARGE](v3){};
\node[U, right of=Z, xshift = 1.2cm, yshift = -2.3cm, font=\LARGE](u3){};

\path (Z) edge [above, line width=0.6mm, draw=black!50] (l1);
\path (Z) edge [above, line width=0.6mm, draw=black!50] (l2);
\path (Z) edge [above, line width=0.6mm, draw=black!50] (l3);

\node [hidden neuron, right of = l1, xshift = 1cm, yshift = -.5cm, font=\LARGE] (h1) {$\sigma$};
\node [hidden neuron, below of = h1, yshift = 0.9cm, font=\LARGE] (h2) {$\sigma$};
\node[annot, below of = h2, yshift = 0.9cm ](dots){$\vdots$};
\node [hidden neuron, below of = dots, yshift = 1cm, font=\LARGE] (h3) {$\sigma$};

\foreach \j in {1,2,3}
    \foreach \i in {1,2,3}
        \path (l\i) edge (h\j);

\node [hidden neuron, right of = h2, yshift = -1cm, xshift = 0cm, font=\LARGE] (out) {$\sigma$};

\foreach \i in {1,2,3}
        \path (h\i) edge (out);
\node [null, right of=out, font=\LARGE](y){};
\path (out) edge [right, line width=0.8mm, draw=black!50] node[text width = 4em, xshift=1.5cm, font=\LARGE]{$\mathbf{y}$}(y);

\node[annot, above of = y, yshift = -1.5cm, xshift = 0.3cm, font=\LARGE]{output};
\node[annot, right of=l1, yshift = 0.7cm, xshift = -0.5cm, font=\LARGE]{$\mathbf{W}_2$};
\node[annot, right of=h1, yshift = -0.8cm, xshift = -0.7cm, font=\LARGE]{$\mathbf{w}_3$};
\end{tikzpicture}
}
\caption{Schematic of the model used for decoding of a multidimensional neural process.\label{fig:LowR}} 
\end{figure}
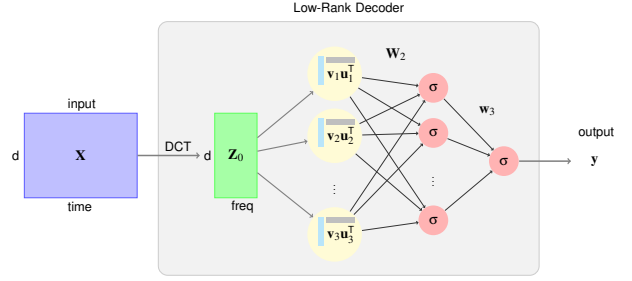

To address this challenge, we propose a novel low-rank neural network structure to reduce
the number of parameters while still enabling
rich nonlinear maps from the response to the stimulus estimate.
Here, we present the model for a regression problem with a scalar target $y$.
However, the same model can be used for classification or multi-target regression
with minor modifications.
Figure \ref{fig:LowR} shows the structure of the model proposed for decoding multidimensional neural processes.
The first stage preprocesses the data by passing each of the $T$ time samples
of the $d$ components through a discrete cosine transform (DCT).  To low-pass filter the
signal, only the first $F$ coefficients in the frequency domain are retained,
hence reducing the dimension from $d\times T$ to $d\times F$.
The low-pass filtering is well-justified assuming that the neural responses to the
stimuli are typically band-limited.

After the low-pass filtering,
the resulting frequency-domain matrix $Z_0 \in \Real^{d \times F}$
is passed through a neural network with two hidden layers and one output layer,
\begin{align}
    Z_{1j} &= u_j^\top Z_0v_j+b_{1j}, \quad j = 1, \dots, h_1,
        \label{eq:layer1} \\
    Z_{2j} &= \sigma(w_{2j}^\top Z_1 + b_{2j}),  \quad
         j= 1, \dots, h_2, \label{eq:layer2} \\
    \hat{y} &= \sigma(w_3^\top Z_2+b_3), \label{eq:layerout}
\end{align}
where $\sigma(t) = 1/(1+e^{-t})$ is the sigmoid function.
The key novel feature of this network is in the first layer \eqref{eq:layer1},
where
each hidden unit $Z_{1j}$ is computed from inner product of the input $Z_0$ with
a rank one matrix $v_ju_j^\top$.   The second hidden layer \eqref{eq:layer2}
and output layer \eqref{eq:layerout} are mostly standard.
The only slightly non-standard component is that, in the output, we have assumed
that the stimuli $y$ is bounded as scaled to a range $y \in (0,1)$ so that we can
use a sigmoid output.

The main motivation of the
rank one structure \eqref{eq:layer1} is to reduce the number of parameters.
A standard fully connected layer would require $Fd+1$ parameters for each hidden unit,
requiring a total of $h_1(Fd+1)$ parameters.  In contrast, the rank one layer \eqref{eq:layer1}
uses only $h_1(F+d+1)$ parameters.  We will see in the results
section that this savings can be considerable.

The low rank structure can be justified, at least heuristically, under the assumption of a 
low rank
structure of the neural responses.  Specifically suppose that the frequency-domain
neural responses, $Z_0$, are approximately given by,
\beq \label{eq:Z0lowrank}
    Z_{0,if} \approx \sum_{k=1}^{h_1} \alpha_k  u_{ki} v_{kf}, \quad i=1,\ldots,d,
        \quad f=1,\ldots,F,
\eeq
where $\alpha=(\alpha_1,\ldots,\alpha_{h_1})$ are some latent variables caused by the stimuli, $y$,
and $u_{ki}$ and $v_{kf}$ are, respectively, the responses of the latent variable $\alpha_k$
over the measurement
channel index $i$ and frequency index $f$.  
Under this assumption, a natural way to estimate the stimulus $y$, is to first estimate
the vector of latent variables $\alpha$ from $Z_0$ and then estimate $y$ from the vector $\alpha$.
Now, we can write \eqref{eq:Z0lowrank} as $Z_0 = G(\alpha)$ where $G(\cdot)$ is a linear map.
The (regularized) least squares estimate for $\alpha$ given $Z_0$ is then given by
$\hat{\alpha} = (G^TG + \gamma I)^{-1}G^T(Z_0)$ for some regularization level $\gamma$.
Due to the separability structure \eqref{eq:Z0lowrank}, it is easily verified that each estimate
$\hat{\alpha}_k$ will be of the form,
\[
    \hat{\alpha}_k = \sum_{j=1}^{h_1} W_{2,kj} u_j^\top Z_0v_j + b_{2k},
\]
for some weights $W_{2,kj}$ and $b_{2k}$.  Hence, the first layers \eqref{eq:layer1} and
\eqref{eq:layer2} of the proposed neural network can be interpreted as recovering the latent variables
under a linear low-rank output model.

\section{Results}
\subsection{$\mu$ECoG data from auditory cortex}
\begin{figure}[t]
\centering
\includegraphics[width = 0.475\textwidth]{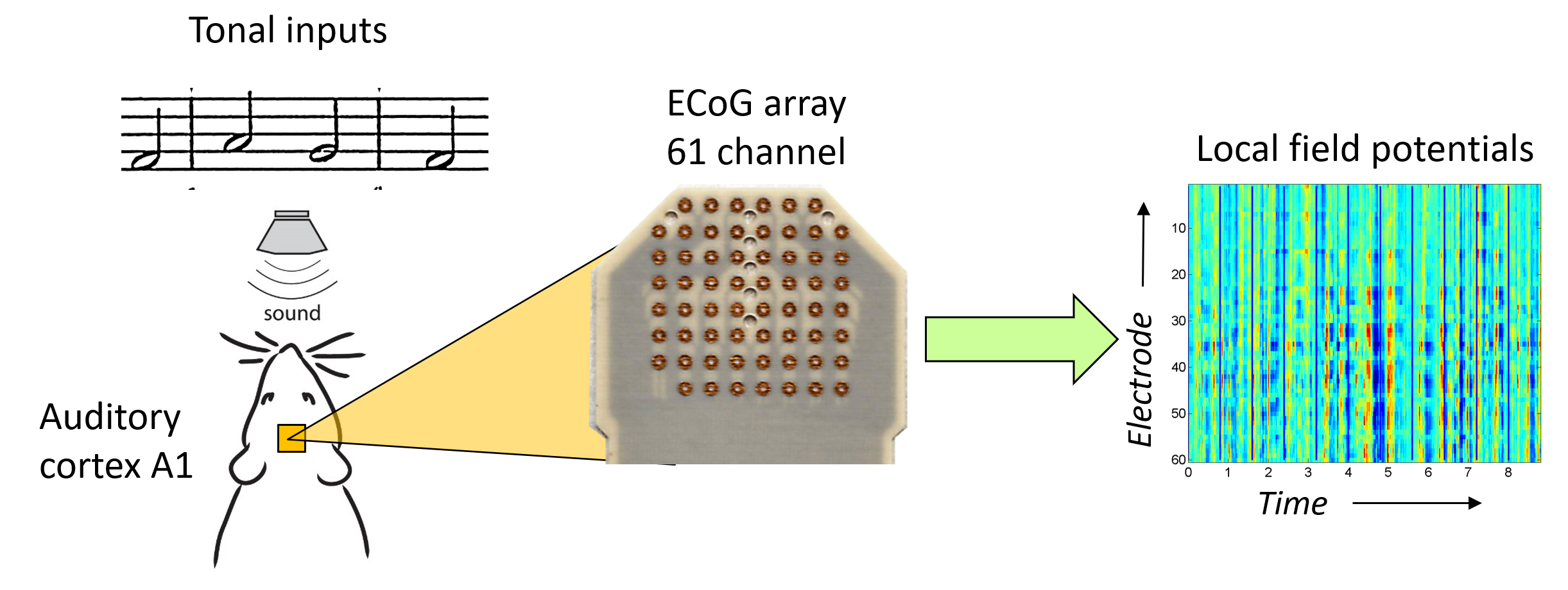}
\caption{Schematic of the experiment setup.}
\label{fig:experiment}
\end{figure}

We evaluate the performance of our model using \textit{in vivo} $\mu$ECoG recordings of A1 area of auditory cortex in moving rodents. Signals are recorded from a high resolution $\mu $ECoG array with electrodes with 420$ \mu m $ spacing. The electrodes were arranged in an 8 x 8 grid where three corner electrodes were omitted \cite{insanally2016low}. In each experiment, single frequency tones with different frequencies are played for $50ms$ every second and the responses are recorded. Figure \ref{fig:experiment} shows the experiment setup and the electrode array.  Recorded signals are then down-sampled to $2000Hz$ for further processing.  There are a total of 390 tones played in each experiment.

\subsection{Decoder performance}
To train our model and test its performance, we generate a dataset $\set{(u^i, X^i)}_{i=1}^{390}$. Each sample $(u^i, X^i)$ consists of the frequency of the stimulus as the input $u^i$ and a $80ms$ window extracted from the signals after the stimulus is applied as the response $X^i$. Since the sampling frequency is $2kHz$, each $X^i$ is a $d\times T$ matrix with $d = 61 $ channels and $T = 160$ time samples. The input frequencies are shifted and rescaled to fall inside the interval $[0.1, 0.9]$. Taking the $F$ point DCT of the signal where $F = 256$, we choose the first 55 frequencies to reduce the dimensionality. We then pass the signal through a low-rank layer with 10 rank-one units. This layer is followed by a Dense layer with 4 hidden units and sigmoid activation. The output layer is a single linear unit with a sigmoid non-linearity which gives us the predicted frequency index. We have used $\ell_2$ regularization with $\lambda = 0.001 $ in learning the weights of both separable and fully connected layers. The model is trained on $66 \%$ of the whole dataset and evaluated on the remaining $34 \%$ as the test set. The goal is to estimate the index of the frequency as a regression problem and R-squared score is used as a measure of closeness of data to the fitted regression model.

\begin{figure}[t]
\centering
\includegraphics[width = 0.5 \textwidth]{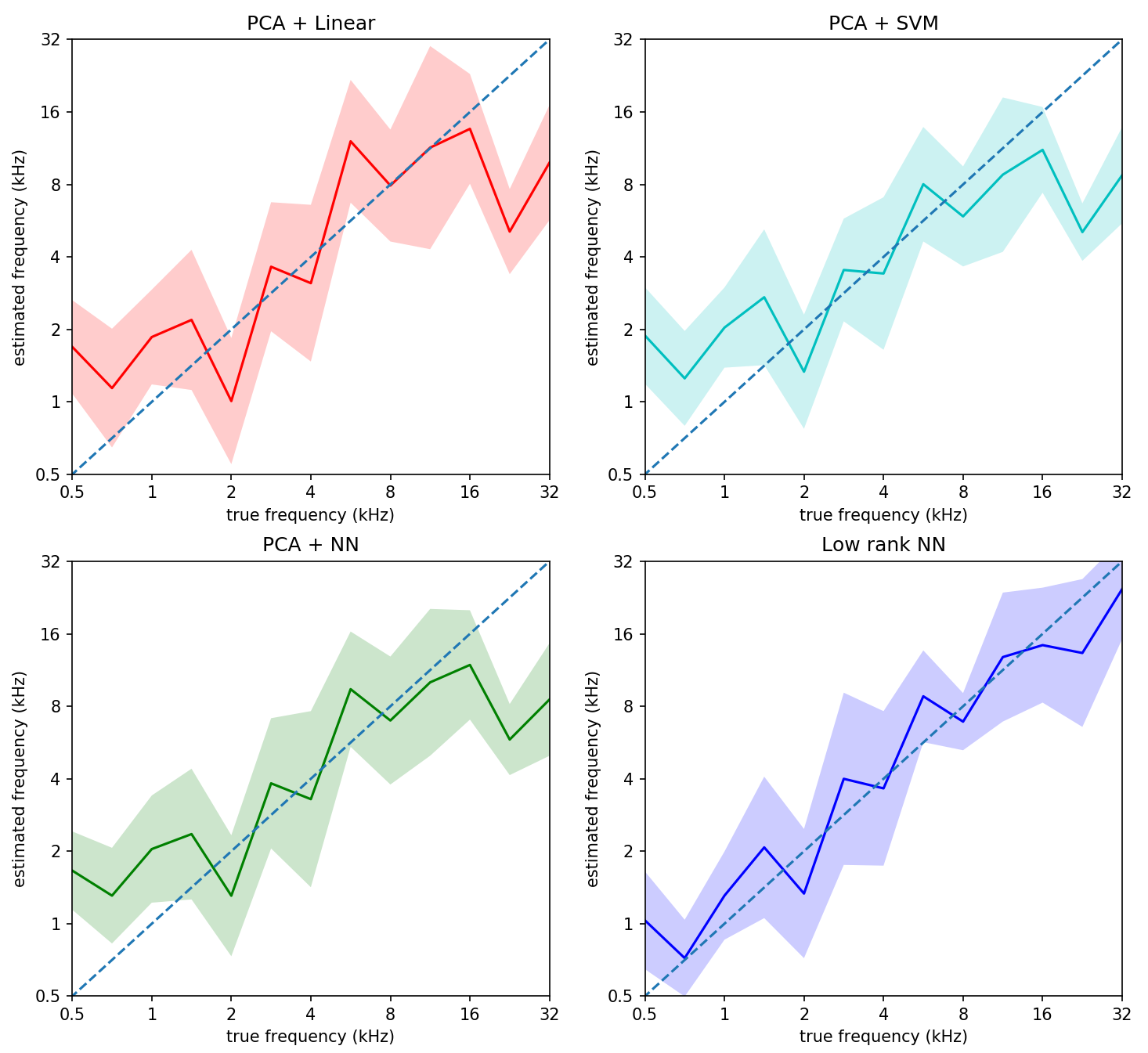}
\caption{Plot of estimated log frequency ($\hat{f}_u$) of the stimulus against the true log frequency ($f_u$) for different models. Dashed line corresponds to the $\hat{f}_u=f_u$ line. The closer the estimated frequency curve to this line is, the performance of the model is better. The proposed low rank neural network model (bottom right) is performing better than the other three models.}
\label{fig:scatter plot}
\end{figure}
We compare the performance of the proposed low-rank neural network with three commonly used models:
\begin{enumerate}
\item \textbf{PCA + linear:} top $p$ principal components of the input are used for linear regression. There are a total of $p+1$ parameters in this model. We use both $\ell_1$ and $\ell_2$ regularizers.
\item \textbf{PCA + SVM:} top $p$ principal components of the input are taken followed by a support vector regressor. There are a total of $p+1$ parameters in this model.
\item \textbf{PCA + NN:} top $p$ principal components of the input are taken followed by a neural network with one hidden layer composed of $n_h$ units. There are a total of $(p+2)n_h +1$ parameters in this model. We use $\ell_2$ regularization for the weights.
\end{enumerate}
For all three models we take top $p=100$ principal components. Cross validation is used to tune the parameters. For SVM, both linear and radial basis function (RBF) kernel were tried and it was found that RBF gives better results.

Figure \ref{fig:scatter plot} shows the performance of all four models in estimating the stimulus frequency on the test dataset. Estimated frequency ($\hat{f}_u$) from each model with one standard deviation error is plotted against the true frequency ($f_u$). The dashed line shows the line $\hat{f}_u=f_u$, corresponding to a perfect model. Therefore, the distance of the prediction curve of each model to this line corresponds to the bias of the estimator and the error shades correspond to the variance of the estimator. The low-rank neural network is closest to the reference line, showing that it is performing better the other models. Table \ref{tab:r_squared} summarizes the performance of each model in estimating the log-frequency of the stimulus in terms of the R-squared metric along with the root-mean-square errors (RMSE).
\begin{table}[t]
\centering
\caption{R-squared score and RMSE of estimating log-frequency of the stimulus for different methods.}
 \begin{tabularx}{0.42\textwidth}{|X|c |c | c||} 
 \hline
 Method & R-squared score& RMSE\\ [0.5ex] 
 \hline\hline
 PCA + Linear & 0.484&  0.179\\ 
 \hline
 PCA + SVM & 0.476  & 0.181\\
\hline
 PCA + NN  & 0.510 &0.174 \\
 \hline
 Low-rank NN & 0.761& 0.121 \\ 
 \hline
\end{tabularx}
\label{tab:r_squared}
\end{table}

\section{Conclusion}
The problem of decoding multidimensional neural responses can be challenging due to high dimensionality of the data. In this work, we presented a neural network model with low-rank structure weights as the first hidden layer which significantly reduces the number of parameters compared to a fully connected network. We tested the model for decoding $\mu$ECoG data recorded from A1 area of auditory cortex of awake rats. We compared the proposed model with some of the most widely used models for decoding neural signals. We showed that our model performs much better in predicting the frequency of the stimulus.
 
\bibliographystyle{apacite}

\setlength{\bibleftmargin}{.125in}
\setlength{\bibindent}{-\bibleftmargin}

\bibliography{ref}

\end{document}